\documentstyle[preprint,revtex]{aps}
\tightenlines
\begin{document}
\draft
\begin{title}
TOWARDS A SINGULARITY-PROOF SCHEME IN NUMERICAL RELATIVITY
\end{title}
\medskip
\centerline{Edward Seidel$^{(1)}$ and Wai-Mo Suen$^{(2)}$}
\medskip
\begin{instit}
$^{(1)}$ National Center for Supercomputing Applications \\
Beckman Institute, 405 N. Matthews Ave., Urbana, IL. 61801
\end{instit}
\smallskip
\begin{instit}
$^{(2)}$ McDonnell Center for the Space Sciences, Department of Physics \\
Washington University, St. Louis, Missouri 63130
\end{instit}

\begin{abstract}

Progress in numerical relativity has been hindered for 30 years because of
the difficulties of avoiding spacetime singularities in numerical evolution.
We propose a scheme which excises a region inside an apparent horizon
containing the singularity.  Two major ingredients of the scheme are the use
of a horizon-locking coordinate and a finite differencing which respects the
causal structure of the spacetime.  Encouraging results of the scheme in
the spherical collapse case are given.

\end{abstract}



\section{INTRODUCTION}\label{intro}

Numerical studies of the Einstein equations promise to deepen dramatically our
knowledge of general relativity, astrophysics, and cosmology.
Shapiro and Teukolsky~\cite{ST86} suggested that the ``Holy Grail of
numerical relativity'' is a code that (i) avoids singularities,
(ii)  handles black holes, (iii)  maintains high accuracy, and
(iv)  is capable of running forever.  These four difficulties in
numerical relativity are closely tied to each other.  Black holes, and within
them spacetime singularities, may be formed after some
evolution, even if none exists initially.  The presence of such objects implies
that
the
dynamical range of the calculation is large, with very
different length and time scales involved.  This makes it
difficult to maintain accuracy or even to keep the code from
crashing in numerical evolution.

The traditional way to deal with this problem is to use the
coordinate degrees of freedom to avoid the ``extreme'' regions.
With ``the many fingers of time'' in relativity, one may evolve other regions
in
space without evolving the region in
which a singularity is about to form.  Many
different types of singularity-avoiding slicings have been
proposed and studied in detail (see, e.g.~\cite{SY78,ES79,BP83}).
This idea of using the freedom in slicing the spacetime to avoid
singularities is ingenious but not perfect.  In the vicinity of the singularity
these
slicings
inevitably contain a region of abrupt change near the horizon,
and a region in which the constant time slices dip back deep into
the past in some sense.  Depending on the details of the choices
of the spacetime coordinates, the code will sooner
or later crash due to these pathological properties of the
slicing.  The problem can appear as the development of spikes in
the spatial metric functions~\cite{SY78}, the steepening of spatial
gradients~\cite{ST86}, ``grid stretching''~\cite{ST86} or large
coordinate shift~\cite{BHS} on the black hole throat, etc.

Hence it is important to investigate other ways to handle
singularities and black holes in numerical relativity.  Cosmic
censorship suggests that in physical situations, singularities
are hidden inside black hole horizons.
With the problematic region of numerical evolution
mostly
inside the horizon, it is tempting to cut away this region from the numerical
calculation by imposing a boundary
condition on or slightly inside the horizon.  To an outside observer no
information is lost since the
region cut away is unobservable.
There is then no singularity to
crash the code, the observable region can be
evolved, the dynamic range is drastically reduced so
accuracy is easier to maintain, and there is in principle no
physical reason that the code cannot run forever.

Using a horizon boundary condition by itself is not a new
idea~\cite{BP83,Unruh,York89,Choptuik92,Thornburgh91}.  However, it is non-
trivial to implement a
horizon boundary condition in dynamical evolution~\cite{Thornburgh91}.  The
boundary
condition to be imposed on a black hole horizon, which is a one
way membrane~\cite{Membrane}, should presumably be some sort of out-going
(into the hole) boundary condition.  However, except for the
case of linear, non-dispersive fields propagating in a flat
spacetime, we are not aware of any satisfactory out-going wave
boundary condition in numerical calculation~\cite{Orzsag}, let alone
for waves in relativity, which can be non-linear, dispersive, and
have tails and other complications~\cite{Waves}.

In this Letter we demonstrate that the idea of a
horizon boundary condition in numerical relativity can indeed be
realized.  In the next section, we discuss the two major
ingredients for our successful implementation of it:  (i)
a ``horizon locking coordinate'' (HLC) system which ties the spatial
coordinates to the spatial geometry,
(ii) a finite differencing scheme that we call causal differencing (CD) which
respects the
causal structure of the spacetime.  CD is
not only essential for the stability of the code using the
HLC, but also eliminates the need of explicitly
imposing boundary conditions on the horizon.  It is, in a sense,
the horizon boundary condition without a boundary condition:
since the horizon is a one way membrane, the quantities on the
horizon can be affected only by quantities outside but not inside
the horizon.  Hence, in CD, all quantities on the boundary can
be updated explicitly in terms of known quantities residing on or
outside the boundary, provided the boundary is inside the horizon.  There is no
need to impose boundary
conditions to account for information not covered by the
numerical evolution.  Such a horizon boundary
condition is particularly
desirable since it is general to all kinds of source terms in the Einstein
equations.

\subsection{Horizon Locking Coordinate}

We use the numerical construction of the Schwarzschild spacetime
to illustrate the idea of locking the horizon.  As shown in~\cite{BHS},
the numerical construction of a Schwarzschild spacetime
is non-trivial.  In all nine differencing schemes and for all
choices of coordinate conditions
studied in ~\cite{BHS}, problems arise after
an evolution in time of about $100 M$ with reasonable choices of grid
parameters.  To facilitate comparison with the results of ~\cite{BHS},
we write the line element in the same form:
\begin{equation}
     ds^2=(-\alpha^2 +\psi ^4  A \beta^2 ) dt^2
  + 2 \psi^4 A \beta dt  d\eta
  + \psi^4 [A d\eta^2
  + B \eta^2 (d\theta^2+sin^2 \theta~d\psi^2 )]
\end{equation}
Such a line
element is easily generalized to one which is suitable for
numerical study of axisymmetric spacetimes~\cite{ABHSS92}, and it includes
both
the radial gauge~\cite{Estabrook73} and the quasi-isotropic or isothermal
gauge~\cite{ST86,Evans86}.

In ~\cite{BHS}, the initial data used in evolving the Schwarzschild
geometry are determined by time symmetry
and conformal flatness~\cite{York86},
that is, the initial slice is an Einstein-Rosen bridge.  They find that the
most
successful code is one
which uses the MacCormack or Brailovskaya differencing scheme, maximal
slicing~\cite{SY78} and zero shift.
This code (denoted by BHS) is
considered to be very accurate, but as for
all codes designed so far to handle black
holes, it will develop difficulties at late times.   In Fig.1,
the dashed line labeled  ``Free Horizon''
gives the coordinate position of the apparent horizon (AH) vs. time.  We see
that
the AH is rapidly
growing in radius.  The solid lines give the value of the radial
metric component $A$ every $t = 10M$.  A spike
is rapidly developing near the
AH, which eventually causes the code to become inaccurate and crash.  The
inaccuracy generated by the sharp spike shows up both in
the violation of the Hamiltonian constraint,
(dashed line with asterisks in Fig. 2) and the drifting of the
mass of the black hole from its initial value 2~\cite{FN2}.

A natural way to avoid this development of a sharp peak is to tie
the coordinate grid points to some
geometric structure.  To achieve this,
(i)We make the coordinate position of the AH
constant in time after it grows
to a certain finite radius.
This determines the
shift $\beta$ at the AH.  (ii)After the horizon is locked,
all grid points are tied to it by
requiring the radial metric function $A(t,\eta)$ to be a constant
in time.  This determines the shift elsewhere.  (iii)  We drop most grid
points from the calculation inside the AH, leaving only a small buffer zone.

The results using the HLC are shown in the
Figs.~\cite{FN3x}.
Fig. 1 shows that after the shift is phased in, the AH is locked and instead of
developing a spike as in BHS or other codes, the radial metric function $A$ is
$O(1)$ everywhere and
absolutely unchanged over time.
In the HLC with the singular region cut away,
{\it there is no need to use any singularity-avoiding time slicing}.
To demonstrate this freedom, we used a lapse that
is constant in time after $t=5M$ (but $\alpha \geq 0.3$ everywhere), when the
AH is securely locked~\cite{FN4}.
Without the sharp spike in the metric function, Fig. 2 shows that outside the
AH
the
Hamiltonian constraint is satisfied to more than an
order of magnitude better than in BHS by $t=100M$, and that
the mass of the black hole remains essentially constant for all time as it
should be in this vacuum case.

\subsection{Causal Differencing}
\medskip

One consequence of introducing a shift vector in the HLC is that inside the AH
the future light cone is tilted inward towards
smaller $\eta$.  This feature is essential for our implementation
of the AH condition.  It allows us to do without a specific
boundary condition for each type of ingoing wave since inside the AH, data
at a particular grid point depend explicitly only on past data from grid points
at
equal or
larger $coordinate$ points.  The existence of a shift calls for a finite
differencing
scheme different from the usual one.  Here we illustrate this with a
simple example.

Consider a scalar field $\phi$ in a 1+1 flat space:
$ ds^2 = -dt'^2 + dx'^2$
and
$\partial^2_{t'} \psi = \partial^2_{x'} \phi$.
Suppose we break it up for numerical evolution as~\cite{firstorder}
$\partial_{t'} \phi = \partial_{x'} \pi$
and
$\partial_{t'} \pi = \partial_{x'} \phi$.
If we introduce a constant shift $\beta$ by the coordinate change $t=t'$ and $x
=
x'-\beta t'$, then the wave equation becomes
$
\partial_t \phi = \partial_x \pi + \beta \partial_x \phi \label{shiftphi}
$
and
$\partial_t \pi = \partial_x \phi + \beta \partial_x \pi. \label{shiftpi}
$
The question is how to write down the finite difference
version of these two equations with a shift.
One might naively use, say, the usual leapfrog scheme:
\begin{equation}
\label{centerleap}
\frac{\phi^{n+1}_j - \phi^{n-1}_j}{2 \Delta t} = \frac{\pi^{n}_{j+1} -
\pi^n_{j-
1}}{2 \Delta x} + \beta \frac{\phi^{n}_{j+1}-\phi^{n}_{j-1}}{2 \Delta x}
\end{equation}
and likewise for the $\partial_t \pi $ equation.  However,
a von Neumann stability analysis shows
that for
any given $C={\Delta t} / {\Delta x}$, the scheme is
unstable for a large enough $\beta$.  For example, numerical experiments show
that for
$C = 1/3 $, the instability quickly takes over
with $\beta C= 0.67$.

To obtain a stable finite differencing,
we do the following:  (i) Return to the unshifted wave equations in the
primed coordinate,
which have untilted light
cones.  The finite differencing can be written as usual.  (ii)
Transform directly the finite difference equation into the unprimed
coordinates.  In leapfrog scheme, this leads to
\begin{equation}
\frac{\phi^{n+1}_{j} - \phi^{n-1}_{j}}{2 \Delta t} =
\frac{\pi^{n}_{j+1+\frac{\Delta x'}{\Delta x}} - \pi^n_{j-
1+\frac{\Delta x'}{\Delta x}}}{2 \Delta x} + \beta^n_j \frac{\phi^{n-
1}_{j+2\frac{\Delta x'}{\Delta x}}-
\phi^{n-1}_{j}}{2 \Delta x'}
\end{equation}
where $\Delta x' = \beta \Delta t$.
Notice that on the RHS, the spatial derivative
is centered at $j+ \beta C$, which is the center
(on the  $n^{th}$ slice) of the causal dependence of the point (n+1,j)
where $\phi$ is to be updated.  This guarantees that the region of causal
dependence is completely
covered whenever $C \leq 1$.

The above procedure of obtaining a finite difference scheme which observes the
causal structure in the presence of a shift vector can be applied to any
difference
scheme, just as for the leapfrog example here.  The underlying idea of CD is
similar to the ``up-wind'' differencing scheme used in
hydrodynamic calculations~\cite{Upwind}~\cite{Schutz92}.  Several variations
on the form of the CD operator will be discussed in a future
paper.

\section {The Apparent Horizon Scheme in Spherical Collapse}

\medskip

In this section we demonstrate these ideas with a spherical collapse code.
A self-gravitating massless
Klein-Gordon scalar field in an initial
time-symmetric gaussian distribution is added to the black hole.
Theoretical and computational details will be presented in a future paper.
Here we summarize results for a typical case.

In Fig. 2b solid (dashed) lines represent the AH mass
$M_{AH}$ as a function of
time for the cases with (without)
the AH scheme.
The ADM mass of the entire spacetime for the case presented
here was $M_{ADM} = 2.62$, while initially  $M_{AH} =
2.08$.  With the gaussian distribution of the matter initially centered at
$\eta = 2.5$ and a width of $\delta \eta = 0.5$,
the hole has absorbed nearly all ingoing radiation by $t=15$.  We see that with
the AH scheme, the
solid line levels off after that time as expected, with a mass of 2.33.
We see no reflection of radiation from the horizon.
The dashed line shows results with BHS where $M_{AH}$
continues to drift upward.
In fact, because of the large spike which develops in the metric
function $A$, the system becomes unstable at about $t = 80M$.  The comparison
of the accuracy~\cite{FN5} between the AH and BHS
schemes is also shown in Fig. 2a.  Again we
see that the violation of the hamiltonian constraint is increasing in the
BHS case, while it is essentially constant over time with the AH scheme,
suggesting that the code is capable of running for a long time.
One final note is that the AH scheme is also potentially much faster
than one using maximal slicing, which involves solving an elliptic equation.

\section{Conclusions}
\medskip

Progress in numerical relativity has been hindered for 30 years because of the
difficulties of avoiding spacetime singularities in the calculations.  We have
presented working examples of how an apparent horizon boundary scheme can
help
circumvent these difficulties in dynamic spacetimes.

There are a number of issues which must be addressed in future work.  For
example, in some codes the constraint equations are solved explicitly for some
components of the extrinsic curvature
or the metric during the
evolution.  These elliptic equations would require boundary conditions which
may be difficult to formulate in an AH scheme.  However, in a free
evolution scheme such issues do not arise.  Also, it is possible for a time
slice to
hit a singularity before an AH is formed (regardless of questions
about Cosmic Censorship)~\cite{SY78}.  This potential problem can be
handled by using a singularity avoiding slicing condition until the horizon is
formed and locked in place, as we have done.  Another potential difficulty is
that
the AH location may jump discontinuously, or
multiple horizons may form.  We believe this problem can be handled by
simply tracking newly formed horizons and phasing in a new AH
condition in place of the old one(s).
Multiple black holes can be handled by a variation of the HLC.
We are beginning to examine these issues
now with both the 2D axisymmetric NCSA black hole code~\cite{ABHSS92}, and
a 3D code using harmonic slicing~\cite{BonaMasso91}, and will report on this
work in future papers.

It is a pleasure to acknowledge discussions with Peter
Anninos, Andrew Abrahams, David Bernstein, Matt Choptuik, David Hobill,
Ian Redmount, Larry Smarr, Jim Stone, Kip Thorne, Lou Wicker, and Clifford
Will.  We are
very grateful to David
Bernstein for providing a copy of his black hole code, on which we based much
of this work.

\bibliographystyle{siam}

\begin{thebibliography}{10}

\bibitem{ST86}
 S. L. Shapiro and S. A. Teukolsky, in {\it Dynamical Spacetimes and
Numerical Relativity},  J.  Centrella, ed.,
Cambridge University Press, Cambridge, (1986).

\bibitem{SY78}
L. Smarr and J. York, Jr., {\it Phys. Rev. D} {\bf 117}, 1945
(1978).

\bibitem{ES79}
D. Eardley, and L. Smarr, {\it Phys. Rev. D}  {\bf 119},
127 (1979).

\bibitem{BP83}
J. Bardeen and T. Piran, {\it Phys. Rep.} {\bf 196}, 205
(1983).

\bibitem{BHS}
D. H. Bernstein, D. W. Hobill and L. L. Smarr, in {\it Frontiers in
Numerical Relativity}, C. Evans, S. Finn, and D. Hobill, eds., Cambridge
University Press, Cambridge, (1989);  D. H. Bernstein, D. W. Hobill, E.
Seidel, and L. L. Smarr, to be published.

\bibitem{Unruh}
W. Unruh, 1984 cited in J. Thornburg,
{\it Class. Quan. Grav.} {\bf 14}, 1119 (1987).

\bibitem{York89}
J. York, in {\it Frontiers in Numerical Relativity}, op. cit.

\bibitem{Choptuik92}
M. Choptuik, private communication.

\bibitem{Thornburgh91}
J. Thornburg, in a talk given at the Toronto Meeting on Numerical Relativity,
May, 1991.

\bibitem{Membrane}
For description of physical properties of black hole horizon as a
one-way membrane, see, {\underline The Black Hole Membrane Paradigm},
edited by K. S. Thorne, R. H. Price, and D. A. Macdonald (Yale University
Press, 1986)

\bibitem{Orzsag}
For some progress in constructing out-going wave boundary
condition for a dispersive wave, see, M. Israeli and S. A. Orzsag, J. Comp.
Phys. 41, 115 (1981).

\bibitem{Waves}
For complications of wave
propagating on a curved background,
see, for example, C. W. Misner, K. S. Thorne, and J. A. Wheeler,
{\em Gravitation } (Freeman, 1973).

\bibitem{ABHSS92}
Abrahams, A., Bernstein, D., Hobill, D., Seidel, E., and Smarr,
L., {\it Phys. Rev.} {\bf
D15}, 1992 (in press); Bernstein, D., Hobill, D., Seidel, E., and Smarr,
L. and J. Towns, to be published.

\bibitem{Estabrook73}
F. Estabrook, H. Wahlquist, S. Christensen, B. Dewitt, L. Smarr,
and E. Tsian, {\it Phys. Rev. D} {\bf 17}, 2814 (1973).

\bibitem{Evans86}
C. Evans, in {\it Dynamical Spacetimes and Numerical Relativity},
op. cit.

\bibitem{York86}
See, for example, J. York in {\it Dynamical Spacetimes and
Numerical Relativity}, op. cit.

\bibitem{FN2}
All results are all obtained with 400 zones and second order spatial
derivatives.  We note that the results obtained with the BHS code are
more accurate if fourth order spatial derivatives or much higher spatial
resolution is used.  However, at late times ($t~100M$)the code will always
develop problems
due to the very steep gradients.

\bibitem{FN3x}
So far we have only used a first order version
of CD discussed in Sec. IIB.  More accurate causal
derivatives, interpolated to the causal center, will be
discussed in a future paper.  At the very inner boundary inside the horizon we
have used both extrapolation and fully one-sided derivatives with excellent
results.

\bibitem{FN4}
However if we continue to use $trK = 0$ throughout the evolution, the results
are slightly more accurate.

\bibitem{firstorder}
The subsequent analysis is essentially the same
if the system is written as $\partial_{t'}
\phi = \pi, \partial_{t'} \pi = \partial^2_{x'} \phi$.

\bibitem{Upwind}
For a detailed study of upwind differencing scheme in relativistic
hydrodynamics, see J. Hawley, L. Smarr, and J. Wilson in {\it Astrophys. J.
Suppl.} {\bf55}, 211, (1984).  Numerical treatment of fluid flow is different
from
that of a shift vector in the sense that there are real physical phenomena
associated with fluids (e.g. shocks), whereas a shift vector just shifts the
coordinates and produces no physical effect.  In the latter case, we can first
turn
off the shift to determine the best differencing scheme which leads to causal
differencing.  In the fluid case, one cannot turn off the fluid flow, as that
completely changes the physics.

\bibitem{Schutz92}
This idea is also very similar to ``Causal Reconnection'', introduced by
M. Alcubierre and B. Schutz, in a recent preprint.  We became aware of this
preprint after our paper was essentially complete.  We thank Richard Matzner
for
bringing this paper to our attention.

\bibitem{FN5}
We note that with
BHS, if we double the resolution, the
error takes off at a later time and the results are closer to the AH results.

\bibitem{BonaMasso91}
C. Bona and J. Masso, {\it Phys. Rev. Lett} {\bf 68}, 1097, (1992).



\end{thebibliography}

\newpage

\figure{The evolution of the radial metric function $A$ and the coordinate
position of the AH are plotted for both the BHS code and the
HLC code.  The dashed line marked ``Free Horizon'' shows the
coordinate position of the AH without the HLC,
while the dashed line is the ``Locked Horizon''.
The solid lines marked ``Free $A$'' show the evolution (in time intervals of
$10M$) of the radial metric function A with no shift, while the solid lines,
which do not change after $5M$, represent the locked case.}

\figure{(a)  For each of the 4 runs we show a measure of the error in the
hamiltonian
constraint as a function of time.  Each point represents
$\sqrt{\sum_{j=j_{ah}}^{j_{max}} G_{tt}(j)^2}$, i.e., the sum of the error
outside
the horizon.  The solid (dashed) lines are obtained with (without) the AH
scheme.  (b) The mass of the AH is shown as a function of time. }

\end{document}